\documentclass[%
 reprint,
 amsmath,amssymb,
]{revtex4-1}

\usepackage{dcolumn}
\usepackage{bm}
\usepackage{mathrsfs}
\usepackage[dvips]{graphicx}
\usepackage[colorlinks=true,urlcolor=black,citecolor=red]{hyperref}
\usepackage{subfigure}
\usepackage{amsmath,amssymb} 
\usepackage[ruled, linesnumbered]{algorithm2e}
\usepackage{setspace}
\usepackage{float}
\usepackage{multirow}
\usepackage{breakurl}

\begin{document}


\title{Naming Game on Networks: \\ \normalsize Let Everyone be Both Speaker and Hearer}

\author{Yuan Gao}
\author{Guanrong Chen}%
\author{Rosa H. M. Chan}%
 \email{rosachan@cityu.edu.hk}
\affiliation{%
 Department of Electronic Engineering, City University of Hong Kong
}%


%

\date{\today}

\begin{abstract}
  To investigate how consensus is reached on a large self-organized peer-to-peer network, we extended the naming game model commonly used in language and communication to Naming Game in Groups (NGG). Differing from other existing naming game models, in NGG, everyone in the population (network) can be both speaker and hearer simultaneously, which resembles in a closer manner to real-life scenarios. Moreover, NGG allows the transmission (communication) of multiple words (opinions) for multiple intra-group consensuses. The communications among indirectly-connected nodes are also enabled in NGG. We simulated and analyzed the consensus process in some typical network topologies, including random-graph networks, small-world networks and scale-free networks, to better understand how global convergence (consensus) could be reached on one common word. The results are interpreted on group negotiation of a peer-to-peer network, which shows that global consensus in the population can be reached more rapidly when more opinions are permitted within each group or when the negotiating groups in the population are larger in size. The novel features and properties introduced by our model have demonstrated its applicability in better investigating general consensus problems on peer-to-peer networks. 
\end{abstract}

\pacs{Valid PACS appear here}
\maketitle


\section{Introduction}
With the rapid development of the Internet, we are well-connected to each other through a peer-to-peer network. A much larger number of new applications can achieve consensus automatically and spontaneously based only on local information and coordination today. For example, new words and abbreviations emerged and were gradually accepted by large populations, such as \emph{bitcoin}, \emph{selfie}, \emph{MOOC}, to name just a few. As a matter of fact, the phenomenon of spontaneous consensus has been extensively discussed in various fields, ranging from linguistics \cite{Loreto07, Steels12, Vylder06}, biology \cite{Rands03}, social sciences \cite{Baronchelli08, Baronchelli10, Castellano09, Fu08, Maity12, Puglisi08, Liu13}, to artificial intelligence \cite{Baronchelli06, Steels96}. The underlying principle of such self-organized consensus has attracted growing research interests in various scientific communities.

Combining complex networks and social dynamics, the naming game (NG) theory provides an effective approach to studying self-organized consensus by mathematically modeling and simulating the consensus processes. Specifically, NG is an interaction-diffusion process on a large-scale network of agents who are trying to reach an agreement on the names of some unknown objects. It has been extensively studied and used for analyzing behavioral consensus problems, such as language evolution \cite{Loreto07, Steels12, Vylder06}, opinion spreading or negotiation \cite{Liu11, Maity12, Yang08}, cultural development \cite{Baronchelli08, Baronchelli10, Puglisi08}, and community formation \cite{Lu09, Xie11, Zhang10}. Recently, NG has been investigated on various complex network models, such as random-graph networks \cite{Baronchelli07, Barrat07, Dall06a, Wang07}, small-world networks \cite{Baronchelli07, Barrat07, Dall06b, Liu11, Liu09, Yang08, Wang07} and scale-free networks \cite{Barrat07, Dall06b, Tang07, Wang07, Yang08}.

\begin{figure}[t]
\centering
\includegraphics[width=0.8\linewidth]{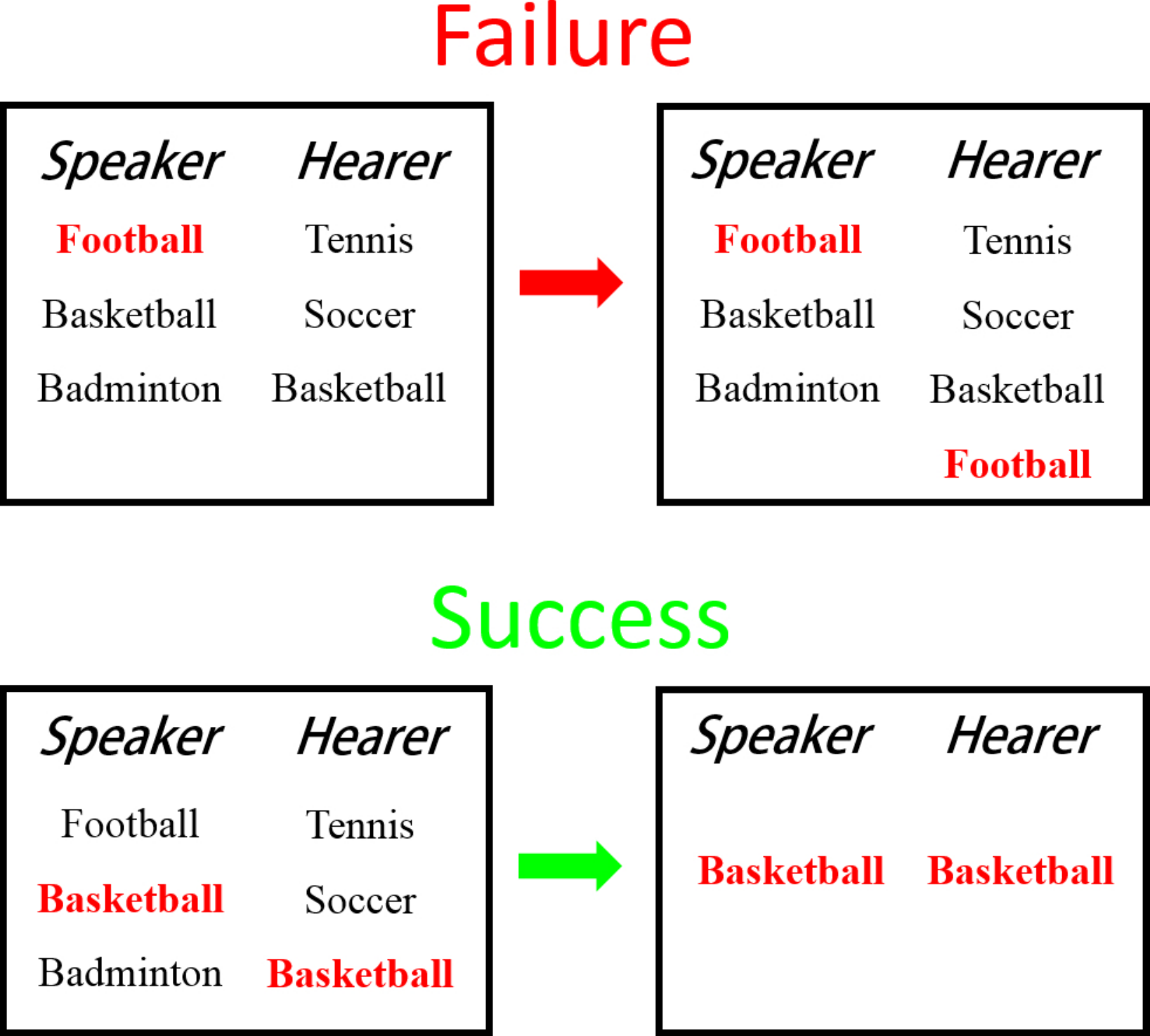}
\caption{(Color online) The interaction of minimal Naming Game. Word in red color is transmitted from the speaker to the hearer. If the hearer did not known the transmitting-word, the conservation fails (top figure), the hearer will add the word to its memory; otherwise, it succeeds (bottom figure), both speaker and hearer will keep only that word.}
\label{fig:minimalNG}
\end{figure}

The minimal NG introduced by Baronchelli \emph{et al}. \cite{Baronchelli06} starts from a population of agents with empty memories, connected in a certain topology. After implementing some simple protocols of game rules, the model facilitates all agents to achieve consensus on the name of an unknown object through conversations among the agents. Specifically, a pair of neighboring agents are chosen, one as speaker and the other as hearer, for conversation. Initially, if the speaker has an empty memory, he/she would generate a new word from a vocabulary and then transmit it to the hearer as the name of the unknown object. But if the speaker already had some words in memory, he/she would randomly choose a word from the memory to tell the hearer. If the transmitting-word also exists in the hearer’s memory, then the two agents reach consensus thereby only that word would be kept by both agents. Whereas if the hearer did not have the transmitting-word in memory, then this conversation fails, so the hearer will learn that word and add it to his/her memory. This process continues until a final convergence to a single word in the whole population, or eventually fails to succeed after a sufficient long time of communications. The interactions of the minimal NG are illustrated in Figure \ref{fig:minimalNG}. 

Thereafter, several variants have been proposed based on the minimal NG model. Wang \emph{et al}. studied the situation where each agent has only a finite memory \cite{Wang07}. A reputation parameter is introduced by Brigatti for a modified failure interaction. That is, if the reputation of the speaker is less than that of the hearer at the step of failure, the speaker invents a new word from a vocabulary but meanwhile decreases its reputation by a fixed amount, while the hearer does nothing \cite{Brigatti08}. On the contrary, Baronchelli  modified the success interaction in which only the speaker (SO-NG) or only the hearer (HO-NG) performs the update of memory \cite{Baronchelli11}. In the model of Maity and colleagues, overhearing can take place by multiple overhearers, which performs the HO-NG at success, and the speaker still interacts with the only one hearer \cite{Maity13}. More recently, naming game with multiple hearers (NGMH) was investigated by Li \emph{et al}., where one speaker interacts with multiple hearers simultaneously \cite{Li13}.

Despite the rapid development of various NG models in the past few years, there is a lack of realistic NG models that can better describe real-life scenarios. To improve this, a new model is proposed in the present article. In the following, the limitations of the existing models and the new model are discussed in detail. Without loss of generality, a self-organized consensus is treated as the result of negotiation on a peer-to-peer network which represents individuals as nodes and their interactions as edges.

First, in retrospect, various social negotiation properties were reported based on NG simulations. For example, Yang \emph{et al}. studied the degree correlation of asymmetric negotiation on both scale-free and small-world networks \cite{Yang08}. They demonstrated that a moderate attempt to choose high-degree agents as speaker would facilitate the fastest global consensus. Liu \emph{et al}. investigated optimal convergence with geography-based negotiation on small-world networks, and found that the fastest convergence could be achieved under a moderate variation on the distances among nodes \cite{Liu11}. However, there are several limitations imposed on these models and their variants: First, the roles of the only speaker and the only hearer (or, one speaker, one hearer and multiple overhearers in \cite{Maity13}; one speaker and multiple hearers in \cite{Li13}) are strictly distinguished. There is only one way to spread a word from the speaker to the hearer. While in a real negotiation scenario, there usually are multiple participants who are peers to each other; therefore, each member is actually both speaker and hearer at the same time. Second, the communication can only take place between directly connected pairs of agents, whereas two indirectly connected agents should be able to communicate through a (short) connected path (via the so-called ``word of mouth'' spreading). Third, in existing models, multiple words are not allowed to be transmitted within a group and multiple intra-group consensuses are not accepted in each step. 

To release the above limitations, herewith we develop a new model, called Naming Game in Groups (NGG). This model generalizes most of the aforementioned models including the most recent NGMH \cite{Li13}. The novel features of NGG are first summarized as follows:
\begin{itemize}
\item The agents of each group are both speaker and hearer simultaneously.
\item Transmission of words between two originally indirectly-connected agents in the same group is allowed to occur through a (short) connected path between them.
\item Multiple words (opinions) are allowed to be spread within each group and multiple intra-group consensuses can take place at each step of the conversation towards global consensus of the whole network. 
\end{itemize}

\section{The New Model}

In the following, the model is described in detail. The NGG model inherits the fundamental structure of the minimal NG \cite{Castellano09}. Basic elements, including network, agent (node), communication (negotiation), speaker, hearer, word (opinion), memory and vocabulary, are preserved. Infinite memory is assumed for all agents, and each agent cannot hear himself/herself (i.e., a network has no self-loops).

Specifically, given a population of agents as a connected network of $M$ nodes discribed by the adjacency matrix $\boldsymbol{A}=[A_{ij}]$, the NGG model is constructed iteratively by group formation, transmitting-words determination and words transmission.

\subsection{Group formation}  \label{sec:group}
A connected sub-network, namely a group $\boldsymbol{G}$ of size $N$ ($N \leq M$), is chosen at random: (i) a node is randomly chosen from the population as the seed, with degree $d_{seed}$; (ii) a number of $min(d_{seed}, N-1)$ neighboring nodes (directly-connected to the seed) are randomly chosen. Thus, a group $\boldsymbol{G}$ is formed, containing $min(d_{seed}+1, N)$ group members in total. Clearly, according to the construction, the maximal path length between any two group members is precisely 2.

\subsection{Transmitting-words determination}  \label{sec:words}
Every group member expresses his/her opinion by saying a word for negotiation. In this scenario, every group member is both speaker and hearer simultaneously, and all the unique words refer to as the set of candidate words, $\boldsymbol{CW}$, for further transmission.

Then, several words are chosen from $\boldsymbol{CW}$ to transmit among both directly- and indirectly-connected group members. For directly-connected pairs, the words can be definitely transmitted as in other existing models; while for an indirectly-connected pair of group members, since the path length between them is 2, the transmission probability is set to 0.5. In order to avoid possible ``gabbling'' in the intra-group negotiations, several transmitting-words are chosen according to their weights. More precisely, each transmitting-word is chosen sequentially with a probability according to how many group members spoke it and how many group members heard it directly. This is reasonable according to the ``plurality rule'' in social negotiations \cite{Conradt09}. And the probability is measured by a weight metric. 

Formally, the \emph{weight} metric in the NGG model is defined on three levels: pair-level, $I_p$; node-level, $I_n$; and word-level, $I_w$. They are further discussed in the following.

1) The pair-level weight $I_p$ is used to determine the transmission between two nodes, $i$ and $j$, for both directly- and indirectly-connected cases:
\begin{equation}
I_p(i,j) = \left\{
\begin{array}{rl}
0 & \text{if } i = j, \\
1 & \text{if } A(i,j) = 1,\\
0.5 & \text{otherwise}.\\
\end{array} \right. \label{Ip}
\end{equation}

2) The node-level weight $I_n(i)$ is an intermediate quantity used to calculate the word-level weight $I_w$ below. For a given node $i$, its $I_n(i)$ is the sum of all pair-level weight between this node and all other group members:
\begin{equation}
I_n(i) = \sum_{j \in \boldsymbol{G}} I_p(i,j). \label{In}
\end{equation}
Note that the purpose of introducing $I_n$ is not to distinguish the roles of the group members, but rather, to allow different weights in the same group depending on the underlying network topology. For instance, if the group is a fully-connected sub-network, $I_n$ for each group members will be identical.  

3) Because different agents may speak the same word, the word-level weight $I_w(w)$ for a word $w$ is defined as the sum of the node-level weights for all the nodes that speak $w$:
\begin{equation}
I_w(w) = \sum_{i \in \mathcal{G}} I_n(i), \text{~ if node $i$ speaks word $w$.} \label{Iw}
\end{equation}

Among all words that have been spoken, a particular $w$ is chosen from $\boldsymbol{CW}$ for transmission according to the following probability:
\begin{equation}
p_w(w) = \frac{I_w(w)} {\sum_{w \in \boldsymbol{CW}} I_w(w)}. \label{pw}
\end{equation}
The denominator in \eqref{pw} is a constant for each $w$. Thus, eventually the chosen probability $p_w(w)$ for $w$ depends only on its $I_w(w)$. Specifically, $p_w(w)$ is determined by how many agents who spoke $w$ (according to \eqref{Iw}) and how many agents who could directly hear it (based on \eqref{Ip} and \eqref{In}).

In this model, $\beta N$ words are allowed for transmission, where $\beta$ is a pre-defined proportion parameter used to determine the number of transmitting-words. The above procedure iterates until a set of $\beta N$ transmitting-words, $\boldsymbol{W}$, are chosen by equation \eqref{pw}. Note that the word(s) with higher weight $I_w(w)$ could be chosen more than once to ``persuade'' the other group members during the negotiation, thus the $\beta N$ transmitting-words of $\boldsymbol{W}$ are not necessarily unique.

\subsection{Words transmission} \label{sec:transmission}

Since each agent may hear several words successively during the group negotiation, each agent is restricted to keep only the first successful word, for a good reason that one should not betray the earlier consensus in an iterative game \cite{Axelrod81}. The transmission order is set as the same order by which the words were picked; namely, word(s) with higher $p_w$ should be selected and transmitted earlier. The words will be transmitted by ``broadcasting'' and ``feedback'' as follows.

Since every agent can be both speaker and hearer, when one agent tells a word to another, the former is referred to as a source node so as to avoid possible confusion. For each transmitting-word $w$ in $\boldsymbol{W}$, according to its selection order, ``broadcasting'' occurs in the group from their source nodes. Whether a group member can hear w from “broadcasting” is determined by a hearing probability $p_h$. Since there could be more than one source node for $w$, a group member will hear $w$ only from the source node ``nearest'' to it, thus
\begin{equation}
p_h(i) = \max_{j} {\{I_p(i,j),\  j \in \boldsymbol{S}\}}, \label{ph}
\end{equation}
where $\boldsymbol{S}$ denotes the set of source nodes for the current transmitting-word. Based on \eqref{Ip} and \eqref{ph}, $p_h(i)$ implies that if node $i$ is directly connected to at least one source node, then it definitely hears $w$; otherwise, it has a probability 0.5 to hear $w$.

After obtaining $p_h$ for all the group members, the set of nodes which heard $w$ can be determined, as $\boldsymbol{H}$. For every node in $\boldsymbol{H}$, if $w$ has already in its memory, then it is successful therefore performs HO-NG \cite{Baronchelli06}, i.e., it keeps only this word $w$ in its memory while dropping all other words thereafter; otherwise unsuccessful, consequently $w$ will be added into its memory.

The ``feedback'' scheme is designed for the successes of source nodes in a probabilistic manner, i.e., the source nodes are likely to be successful if many other group members agree with them. Specifically, the success probability of each source node is $n_{succ}/N$, where $n_{succ}$ is the number of nodes succeeded to the current transmitting-word $w$. Note that if there are multiple source nodes, one can also be successful if it heard $w$ ``broadcasted'' by other source nodes. Thus, the ``feedback'' scheme only takes effect on the unsuccessful source nodes after ``broadcasting''.

The above procedure of words transmission repeats, until $\beta N$ words in $\mathcal{W}$ have been transmitted.

\subsection{Global convergence and stopping criterion}
The above-described algorithm iterates on the whole network following the description in Subsections \ref{sec:group}--\ref{sec:transmission} above. Recall that the groups are chosen at random from time to time (Subsections \ref{sec:group}), every group member can only speak words from its memory if its memory is not empty (Subsections \ref{sec:words}), and the intra-group success leads to eliminating non-consent words from group members' memories (Subsections \ref{sec:transmission}). Thus, the iteration of this procedure will eventually lead to global convergence to a common word. Then, the iteration stops.

\subsection{Comparison with the NGMH model}

It is remarked that the proposed NGG model will reduce to the NGMH model \cite{Li13} by two simplifications: (i) letting only the seed node speak a word, i.e., the seed node is the only speaker, and there is only one word to be transmitted during each group conversation, and (ii) using a deterministic function $\lfloor n_{succ}/N \rfloor$ to determine the success of the unique speaker.
\begin{table*}[t]
\fontsize{8pt}{0.8\baselineskip}\selectfont
\begin{ruledtabular}
\begin{tabular*}{\textwidth}{@{\extracolsep{\fill}} l ccccc}
Network & $\#$Nodes   & $<D>$ & $<PL>$ & $CC$ \\
\hline
RG with $P = 0.03$ $(RG-0.03)$ & 1000 & 30.0 & 2.3643 & 0.0294\\
RG with $P = 0.05$ $(RG-0.05)$ & 1000 & 49.9 & 2.0285 & 0.0502\\
RG with $P = 0.1$ $(RG-0.1)$ & 1000 & 99.9 & 1.9000 & 0.1007\\
SW with $K = 20$ and $RP = 0.1$ $(WS-20-0.1)$ & 1000 & 40.0 & 2.6281 & 0.5366\\
SW with $K = 20$ and $RP = 0.2$ $(WS-20-0.2)$ & 1000 & 40.0 & 2.4651 & 0.3837\\
SW with $K = 20$ and $RP = 0.3$ $(WS-20-0.3)$ & 1000 & 40.0 & 2.3517 & 0.2661\\
SF with $n_0 = 25$ and $e = 25$ $(BA-25)$ & 1000 & 46.9 & 2.0727 & 0.1081\\
SF with $n_0 = 51$ and $e = 50$ $(BA-50)$ & 1000 & 89.7 & 1.9133 & 0.1681\\
SF with $n_0 = 76$ and $e = 75$ $(BA-75)$ & 1000 & 121.6 & 1.8700 & 0.1906\\
\end{tabular*}
\end{ruledtabular}
\caption{Network details in simulations, where $P$ is the connecting probability of random-graph network (RG); the small-world network (SW) is initialized by a ring-shaped network with $2K$ degree, and $RP$ is the rewiring probability; the scale-free network (SF) is generated by $n_0$ initial nodes and one node with $e$ edges to existed nodes is added in each step. $<D>$ is average degree, $<PL>$ is average path length and $<CC>$ is average clustering coefficient.}
\label{table:network}
\end{table*}

\begin{figure*}[t]
\centering
\includegraphics[width=0.8\linewidth]{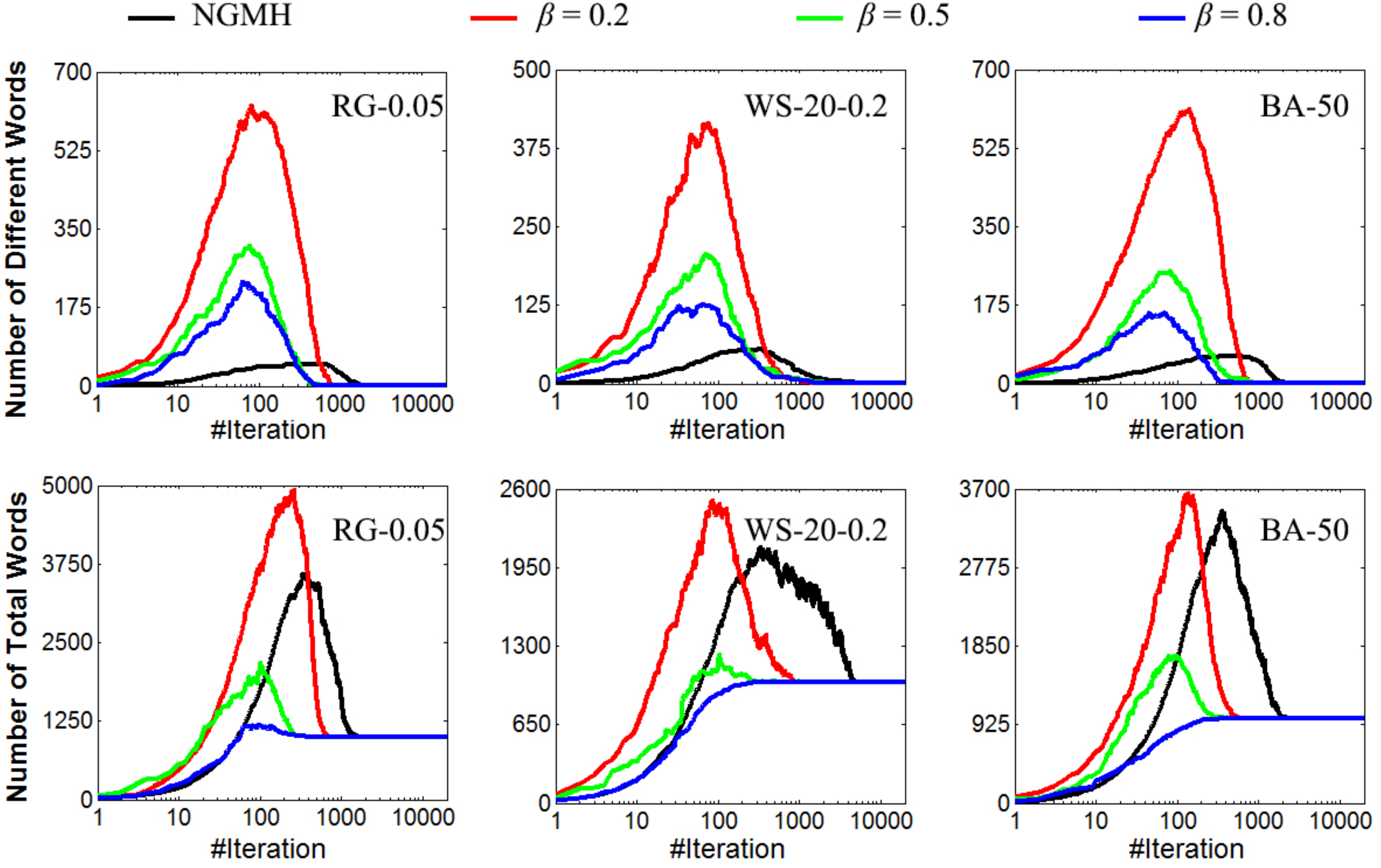}
\caption{(Color online) The convergence of the NGG model for group size $N$ equal to 20. Both Number of Different Words vs. \#Iteration and Number of Total Words vs. \#Iteration are investigated. Three different $\beta$ as well as NGMH are compared. The tested network configurations are $RG-0.05$, $WS-20-0.2$ and $BA-50$.}
\label{fig:instance}
\end{figure*}

\section{Results}
Numerical simulations carried out on NGG are reported here, so as to further investigate the properties and effects of the model parameters $\mathbf{\theta}=\{N, \beta \}$. The relationships between $\mathbf{\theta}$ and the three convergence metrics, $\mathcal{M}=\{N_{total\_max}, N_{diff\_max}, N_{iter\_cvg}\}$, are examined on three representative network models, i.e., random-graph (RG), small-world (SW) and scale-free (SF) networks, of the same size with different configurations. $N_{total\_max}$ is the number of maximal total words appearing in the whole network throughout the iteration process, $N_{diff\_max}$ is the number of maximal different words appearing likewise, and $N_{iter\_cvg}$ is the number of iterations to achieve convergence.

This section is organized as follows. The simulation setup is described firstly. Then, the convergence and simulation results of NGMH and NGG are compared. After that, the NGG model is studied. Both model parameters $\mathbf{\theta}$ and network topologies are analyzed according to their influences on the local success (quantified by the Success Ratio) and the global success, so as to provide an intermediate metric to explicitly interpret the effect due to alteration of $\mathcal{M}$. Thereafter, each convergence metric of $\mathcal{M}$ is discussed based on the local success and the global success, explicitly but in separate subsections.

\subsection{Simulation setup}
In simulations, three different network models with $M = 1000$ nodes are used, and for each model three different configurations are generated following the standard mechanisms \cite{Erd59, Watts98, Barab99}. Here, RG and SF networks with different average degrees ($<D>$) are investigated; as for Small-World networks (SW), only the results of different rewiring probabilities ($RP$) are presented and analyzed here for brevity, since $RP$ alter only the clustering coefficient value ($<CC>$) which governs the ``global connectivity''. The results of WS networks initialized by different connected neighbors $K$ are illustrated in Table S1, Figures S1 and S2 in the Supplementary Material \cite{Supp}. Basic configuration properties of the tested networks are summarized in Table \ref{table:network}. 

To examine the effect of $\mathbf{\theta}$ when $N$ is larger or smaller than $<D>$, three network models with moderate $<D>$ (or moderate $<CC>$, i.e. $RG-0.05$, $WS-20-0.2$ and $BA-50$) are simulated for $N = \{10, 20, 50, 100\}$, respectively, where two of $N$s are smaller than their $<D>$, while the other two are larger. The other six network models of Table \ref{table:network} (i.e., $RG-0.03$, $RG-0.1$, $WS-20-0.1$, $WS-20-0.3$, $BA-25$ and $BA-75$) with $N = 20$ are also simulated, to analyze the effect of different network configurations. All numerical results presented below are obtained by averaging 20 simulations.

\begin{figure*}[t]
\centering
\includegraphics[width=0.8\linewidth]{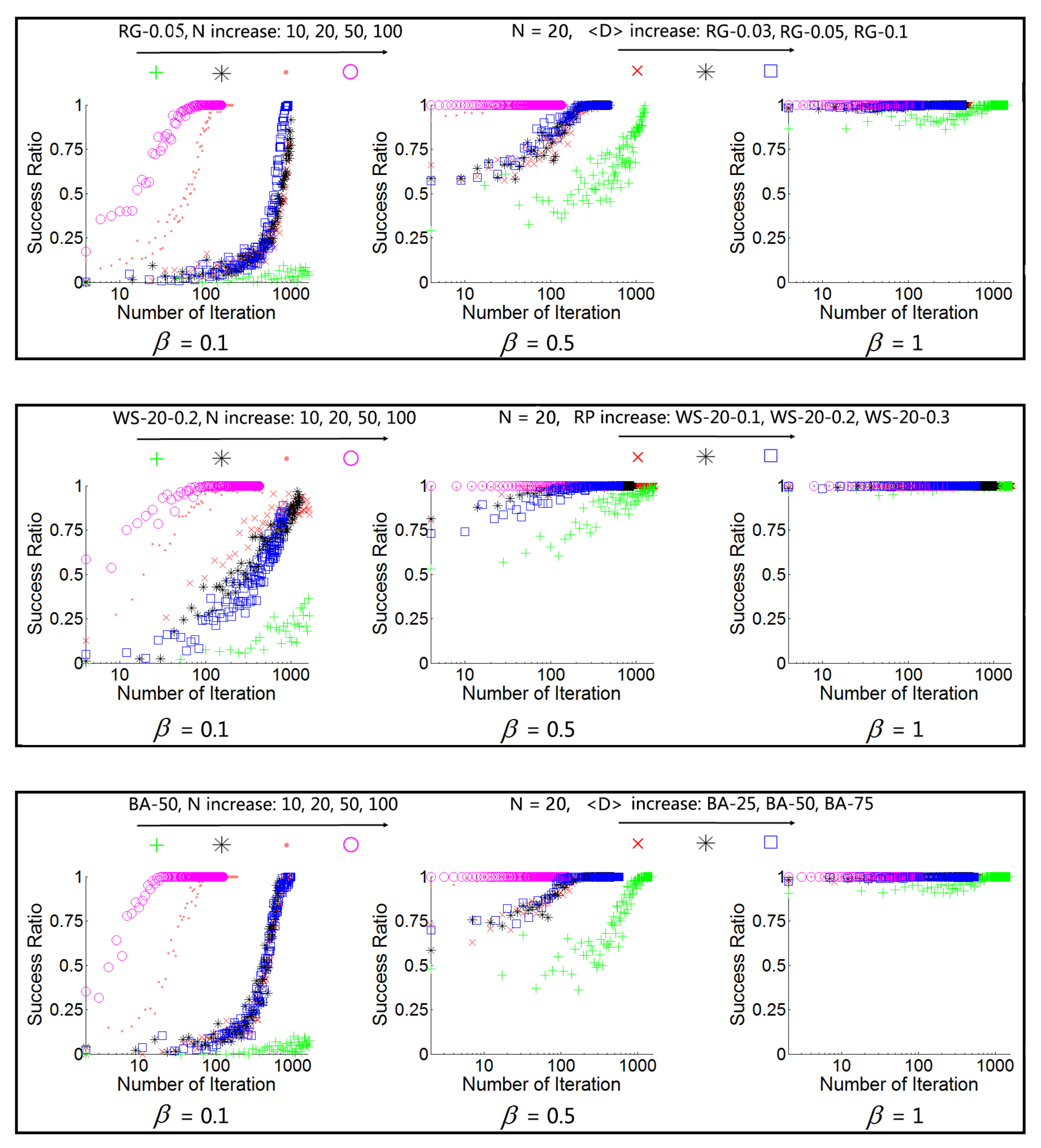}
\caption{(Color online) $SR$ for each iteration with $\beta$ equal to 0.1, 0.5, and 1. Top Row: the $SR$ for RG networks of various group size $N$, i.e., $RG-0.03$ with $N = 20$, $RG-0.05$ with $N = \{10, 20, 50, 100\}$, $RG-0.1$ with $N = 20$. Second Row: the $SR$ for WS networks of various group size $N$, i.e., $WS-20-0.1$ with $N = 20$, $WS-20-0.2$ with $N = \{10, 20, 50, 100\}$, $WS-20-0.3$ with $N = 20$. Bottom Row: the $SR$ for BA networks of various group size $N$, i.e., $BA-25$ with $N = 20$, $BA-50$ with $N = \{10, 20, 50, 100\}$, $BA-75$ with $N = 20$.}
\label{fig:SuccessRate}
\end{figure*}

\subsection{Convergence of NGG}
\begin{figure*}[t]
\centering
\includegraphics[width=0.8\linewidth]{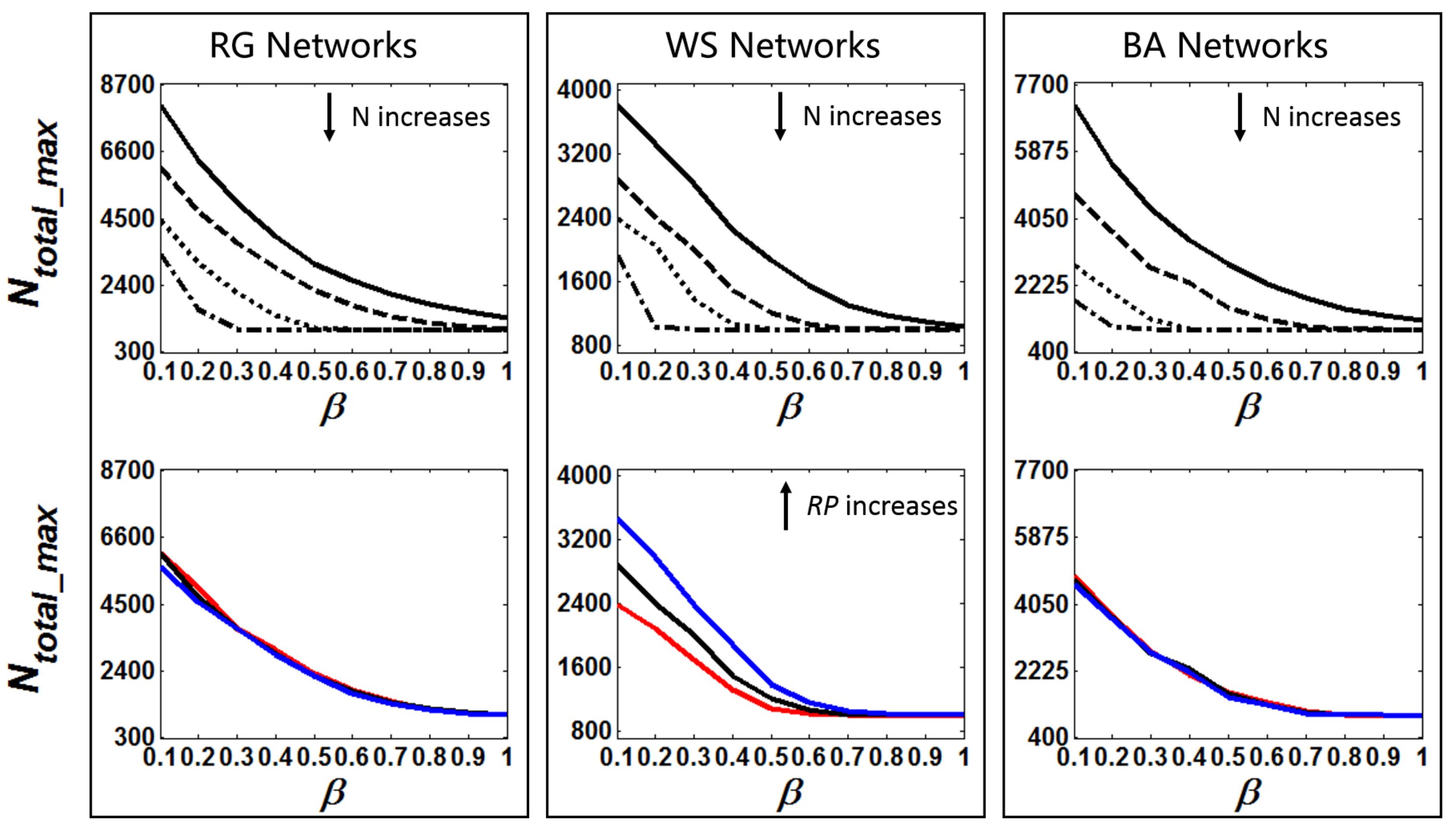}
\caption{(Color online) The relationship between $N_{total\_max}$ and model parameter $\mathbf{\theta}$ in different network configurations. Columns from left to right indicate the simulation results for RG, WS and BA networks. The Top Row is the comparison of different group size $N$ for same network instance, i.e. $RG-0.05$ for RG, $WS-20-0.2$ for WS and $BA-50$ for BA. The group size $N$ from top to bottom of each subfigure in the Top Row is 10, 20, 50, 100, respectively. The Bottom Row is the results of different network instances with same $N = 20$. The network instances (in the order of RED, BLACK, BLUE) are $RG-0.03$, $RG-0.05$, $RG-0.1$ for RG, $WS-20-0.1$, $WS-20-0.2$, $WS-20-0.3$ for WS and $BA-25$, $BA-50$, $BA-75$ for BA.}
\label{fig:NTotalMax}
\end{figure*}

The number of total words, $N_{total}$, and the number of different words, $N_{diff}$, in each iteration are used to investigate the convergence. To be concise, only the results of $N = 20$ with $\beta = {0.2, 0.5, 0.8}$ for $RG-0.05$, $WS-20-0.2$ and $BA-50$ are shown in Figure \ref{fig:instance}; more results can be found in Figure S3 - S5 \cite{Supp}. The NGMH results with same parameters are partly illustrated in Figure \ref{fig:instance} for comparison. For more comparison between NGMH and NGG, please refer to Figures \ref{fig:NTotalMax} - \ref{fig:NIterCvg} and Table S2 \cite{Supp}.

Figure \ref{fig:instance} indicates that the NGG model does converge, since $N_{diff}$ is stabilized at 1 and $N_{total}$ converges to the network size, 1000. Compared with the NGMH model, more different words (i.e., larger $N_{diff\_max}$) can be processed by the NGG model. Because each group member speaks a word in NGG, more different words are generated randomly from the vocabulary at the beginning when the memories of the group members are empty. Moreover, since multiple words are allowed for transmission within each group, the NGG model generates more intra-group consensuses. These which finally lead to faster convergence (i.e., smaller $N_{iter\_cvg}$). As to $N_{total\_max}$, more words are generated but also more words are eliminated (i.e., more intra-group consensuses) in the NGG model. It is found that $N_{total\_max}$ is determined by $\beta$ when other parameters are fixed, as shown in the bottom row of Figure \ref{fig:instance}. The influence of $\beta$ on $N_{total\_max}$ will be further discussed later.

\subsection{Local Success and Global Success}
In order to conduct an explicit analysis on the alteration of convergence metrics $\mathcal{M}$, both local success and global success have been investigated as intermediate metrics. Local success represents the intra-group success during group negotiation, while global success is the success when the whole network converges to one common word.

An parameter named Success Ratio is introduced to represent the degree of local success: $SR = \frac{\#\{Successful\text{ }Group\text{ }Members\}}{\#\{Total\text{ }Group\text{ }Members\}}$. The corresponding $SR$ of selected network configurations is illustrated in Figure \ref{fig:SuccessRate}. In the following subsections, it can be observed that all three metrics of $\mathcal{M}$ are changed monotonously with $\beta$, thus only the results of $\beta = \{0.1, 0.5, 1\}$ are shown here for brevity, while more results can be found in Figures S6 - S8 \cite{Supp}.

From Figure \ref{fig:SuccessRate}, it can be observed that: (i) $SR$ increases with a larger $N$, because the number of transmitting-words is larger by a larger $N$. Thus, words with higher weights are likely to be selected more frequently to facilitate a higher $SR$. (ii) The $SR$ also increases with a larger $\beta$ for the same reason. (iii) Similarly, $SR$ will be saturated at 100$\%$ even during initial iterations if $\beta N$ is sufficient large (e.g., the purple circle in the middle column in Figure \ref{fig:SuccessRate}. (iv) Different $<D>$ values of RG and BA do not influence $SR$ much; and a lower $RP$ (higher $<CC>$) of a WS network leads to a higher $SR$, and a higher $<CC>$ of WS network represents a more ``locally connected'' network. Thus, most of group members likely communicated with each other before the current negotiation takes place, so they are likely to agree with each other.

Differing from the local success, global success is mainly determined by the overall topology of the network, specifically, the ``global connectivity''. Thus, it is affected by the $<CC>$ of the network, i.e., larger $<CC>$ leads to harder in achieving global success.

\subsection{Relationship between $\mathbf{\theta}$ and $N_{total\_max}$}
Let $N_{total\_max}$ be the number of maximal total words that can be processed during iterations (e.g. the maximal value of each curve in Top Row, Figure \ref{fig:instance}). Since the number of total words, $N_{total}$, will keep increasing due to ``broadcasting'' until successes that eliminate more words are achieved. Thus, the earlier and the more intra-group successes are achieved, the smaller the value of $N_{total\_max}$ will be. Specifically, $N_{total}$ increases roughly by $\beta N^2$ words after each iteration without considering the eliminations; on the contrary, more and more $N_{total}$ will be eliminated due to the increase of $SR$ during the iterations, as shown in Figure \ref{fig:SuccessRate}. Here, $N_{total\_max}$ is achieved in mid-iterations when the increased $N_{total}$ is equal to the decreased $N_{total}$. Note that although $SR$ will keep increasing during the iterations, higher $SR$ in late iterations does not affect the value of $N_{total\_max}$. Eventually, $N_{total\_max}$ is inversely proportional to the $SR$ of mid-iterations eventually. Please refer to Figures S9 - S11 for numerical evidences of the above analysis \cite{Supp}.

The simulation results for $N_{total\_max}$ according to different $\mathbf{\theta}$ are illustrated in Figure \ref{fig:NTotalMax}: (i) $N_{total\_max}$ decreases drastically as $\beta$ increases until converging to the network size, 1000, because higher $\beta$ leads to higher $SR$. (ii) Similarly, $N_{total\_max}$ decreases as $N$ increases. (iii) For WS networks, $N_{total\_max}$ increases significantly when the $RP$ increases, whereas such decreasing is not significant in RG and BA networks with different $<D>$. This phenomenon is also consistent with the above analysis on $SR$.

\subsection{Relationship between $\mathbf{\theta}$ and $N_{diff\_max}$}
Let $N_{diff\_max}$ be the maximal number of different words that can be processed during the iterations (e.g., the maximal value of each curve in Bottom Row, Figure \ref{fig:instance}). At the beginning when every node had an empty memory, the number of different words, $N_{diff}$, increases due to the group negotiations. Meanwhile, $N_{diff}$ decreases due to the elimination of (different) words by successful conversations. Since $N_{diff}$ will also be achieved at mid-iterations, ultimately $N_{diff\_max}$ is inversely proportional to the $SR$ of mid-iterations. Please refer to Figures S9 - S11 for numerical evidences of the above analysis \cite{Supp}.

\begin{figure*}[t]
\centering
\includegraphics[width=0.8\linewidth]{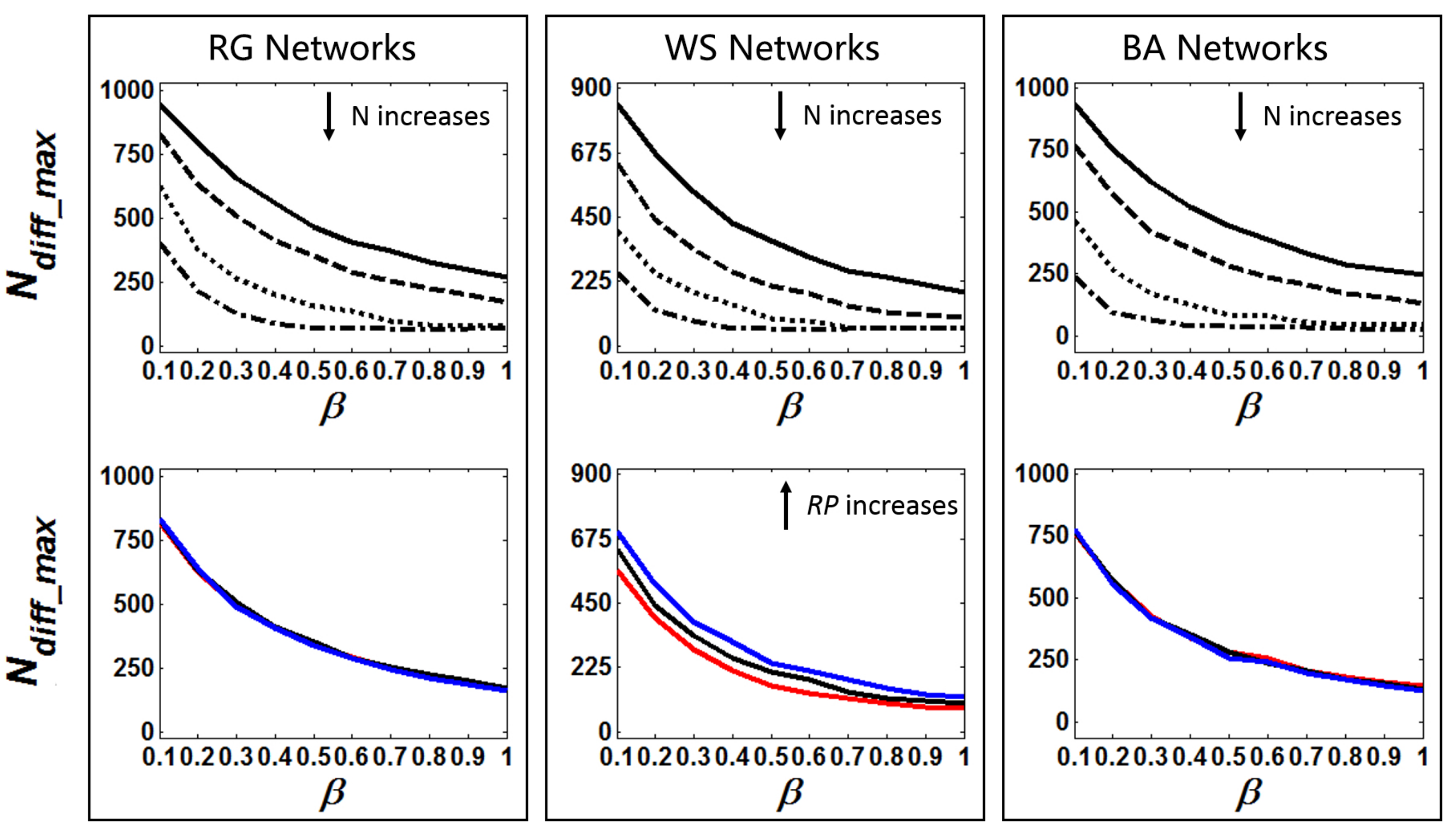}
\caption{(Color online) The relationship between $N_{diff\_max}$ and model parameter $\mathbf{\theta}$ in different network configurations. Other parameters are same as Figure \ref{fig:NTotalMax}.}
\label{fig:NDiffMax}
\end{figure*}

\begin{figure*}[t]
\centering
\includegraphics[width=0.8\linewidth]{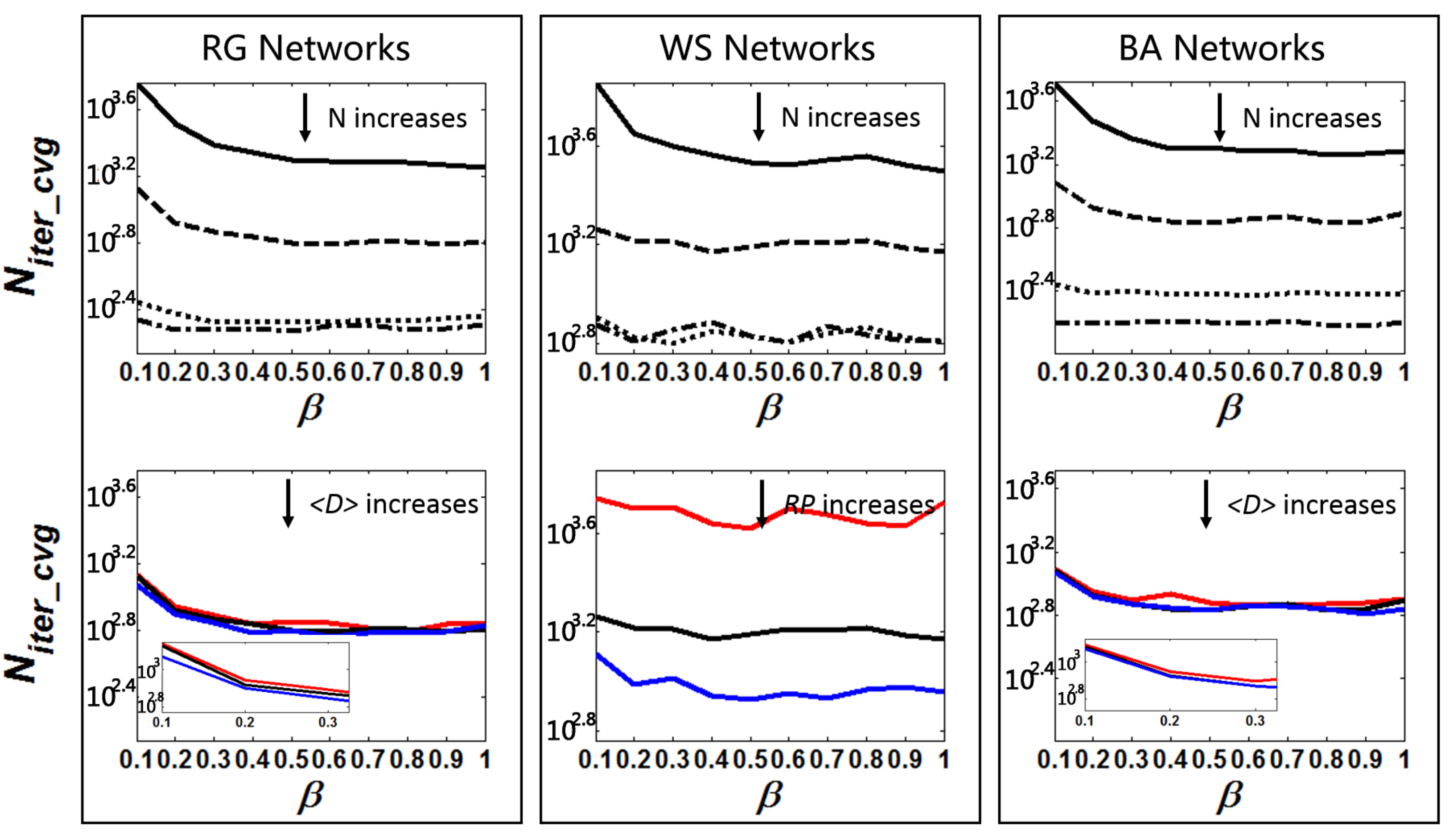}
\caption{(Color online) The relationship between $N_{iter\_cvg}$ and model parameter $\mathbf{\theta}$ in different network configurations (log-scale in Y-axis). Other parameters are same as Figure \ref{fig:NTotalMax}.}
\label{fig:NIterCvg}
\end{figure*}

Since both $N_{total\_max}$ and $N_{diff\_max}$ are inversely proportional to the $SR$, it is expected that the alteration tendency of them should be similar. Actually, the simulation results shown in Figure \ref{fig:NDiffMax} are consistent with one’s expectation: (i) $N_{diff\_max}$ decreases as $\beta$ increases, since larger $\beta$ leads to higher $SR$. (ii) $N_{diff\_max}$ decreases as $N$ increases, because $N$ is also proportional to the $SR$. (iii) According to the performed simulations and analysis on $SR$, $N_{diff\_max}$ increases significantly when $RP$ increases while all $\mathbf{\theta}$ are fixed in WS networks, whereas $N_{diff\_max}$ seems to be stable in both RG and BA networks when different $<D>$ values are used.

\subsection{Relationship between $\mathbf{\theta}$ and $N_{iter\_cvg}$}
Differing from $N_{total\_max}$ and $N_{diff\_max}$, the number of iterations to convergence, $N_{iter\_cvg}$, is determined by the global consensus. Firstly, $N_{iter\_cvg}$ is highly impacted by the network topology, e.g. the global consensus is hard to achieve in WS networks with large $<CC>$, because the common words resulted from intra-group consensuses are not easy to spread out far away. While if $<CC>$ is small, $N_{iter\_cvg}$ will be relatively small. In this case, $N_{iter\_cvg}$ is also affected by intra-group consensus, i.e., $SR$. In summary, $N_{iter\_cvg}$ will be smaller if intra-group consensus can be (achieved and) spread out easier. in other words, $N_{iter\_cvg}$ is proportional to the $<CC>$ and inversely proportional to the $SR$.

Figure \ref{fig:NIterCvg} shows the simulation results on $N_{iter\_cvg}$. It can be observed that: (i) generally, $N_{iter\_cvg}$ decreases as $\beta$ increases, especially when $\beta$ is small. This is because, when $\beta$ is small, increasing $\beta$ leads to a drastic increase in $SR$, as shown in Figure \ref{fig:SuccessRate}. (ii) $N_{iter\_cvg}$ increases as $N$ decreases. Firstly, with a larger $N$, a higher value of $SR$ can be achieved. Also, a larger N leads a larger number of group members to succeed towards a common word thereby facilitating a faster global convergence. (iii) For WS networks, increase in $RP$ leads to decrease in $N_{iter\_cvg}$, because a larger $RP$ results in a smaller $<CC>$, which means the networks are more ``globally connected'' and the intra-group consented words can be easily spread out. (iv) For homogenous networks (RG and WS) with the same configurations, $N_{iter\_cvg}$ decreases insignificantly when $N$ is further increasing after it exceeds $<D>$ (Row 1, Columns 1 and 2 of Figure \ref{fig:NIterCvg}). Because the expected real group size of a homogenous network is $<D> + 1$, further increasing $N$ will not increase the real group size. While the decrease in $N_{iter\_cvg}$ is still significant when $N$ is further increasing after it exceeds $<D>$ in heterozygous networks (BA) (Row 1, Columns 3 of Figure \ref{fig:NIterCvg}). This is due to the fact that increasing $N$ will actually increase the real group size since hub nodes could be selected as seeds to form groups.



\section{Discussion}
In this work, a novel NGG model with a group structure has been developed. Differing from all other models, in NGG, (i) every node is both speaker and hearer; (ii) conversations can take place among indirectly-connected nodes within a group; (iii) multiple words are allowed to transmit in each group towards multiple intra-group consensuses.

Compared to the recent NGMH model \cite{Li13}, the NGG model accelerated faster towards global convergence on a comparable setting with the same group size, because multiple intra-group transmission words, allowed in NGG, lead to more intra-group consensuses. In fact, the NGMH model is a special case of the NGG model when the seed node (to construct group) is the unique speaker and the success of this unique speaker is determined by a deterministic function rather than a probability.

We have analyzed the key parameters introduced into the NGG model, so as to identify and evaluate their effects on the performances over different network topologies. The results show that, in general, the alteration of the group size and the allowable number of transmitting-words have effects in inverse proportion on the maximal number of words, the maximal number of different words, and also the convergence time. 

Some discoveries from the simulations can be interpreted by group negotiation/communication on a peer-to-peer network. In such a network, a large population of peer-to-peer agents negotiates through some small groups (subsets) iteratively so as to achieve global consensus of the whole population. The NGG model allows every agent to be both speaker and hearer simultaneously. Moreover, multiple intra-group consensuses can be described and stimulated by the NGG model. In addition, all the agents in a same group can communicate with each other through a shortest path, even if they are not located side by side (i.e., they are only indirectly-connected). These properties of the NGG model are more applicable for stimulating peer-to-peer self-organized consensus problems. In the considered setting, $N_{total\_max}$, $N_{diff\_max}$, and $N_{iter\_cvg}$ represent the maximal number of words (opinions), maximal number of different words (opinions) and the time needed to achieve global consensus, respectively. Interestingly, it is observed that, generally, the more transmitting words or the larger negotiating group size is allowed, the smaller the maximal number of words and maximal number of different words are, and finally the faster global convergence will be. Additionally, the results suggest that small cliques are harmful to global consensus, e.g., it is difficult to reach consensus in WS networks with large $<CC>$.

Other features of the NGG model comprise many real-world phenomena that involve group negotiations (communications) on peer-to-peer networks. For example, there is a phenomenon that the most successful commercial films made in Hollywood are accepted by people all over the world and earned main box office receipts overseas \cite{BOM14}. By considering the underlying communication network behind such a phenomenon as a worldwide peer-to-peer network, the NGG model is more suitable for describing such related global culture dissemination problems. In short, the NGG model developed in this study not only extends the basic features of naming game, but also sheds lights onto feasible models and approaches that are more applicable for investigating general self-organized consensus problems on peer-to-peer communication networks.

\begin{algorithm*}
\SetNlSkip{0.5em}
\SetInd{0.5em}{1em}
\caption{Naming Game in Groups}
\label{NGG}
Set parameter values for $M$, $N$, $\beta$.  \\
Initialize $\boldsymbol{V}$, $\boldsymbol{A}$ with $M$ nodes, initialize the memory of each node to be null. \\
\Repeat{\rm all nodes keep only one same word}{
 Randomly choose a seed $s$ with degree $d_{seed}$, and randomly choose $\min(d_{seed}, N - 1)$ nodes from $\boldsymbol{A}$ connected to $s$, so as to form a group with $\min(d_{seed} + 1, N)$ group members, $\boldsymbol{G}$. \\
 Every group member speaks a word, forming a UNIQUE set of candidate words, $\boldsymbol{CW}$.\\
 For every word $w \in \boldsymbol{CW}$, calculate $p_w$ by equations (2), (3), (4) and (5). \\
 Based on the $p_w$, choose $\beta N$ words from $\boldsymbol{CW}$, forming transmitting-words set $\boldsymbol{W}$ \\
 Initialize a set of unsuccessful group members, $\boldsymbol{U} = \boldsymbol{G}$. \\
 \ForEach{{\rm word $w \in \boldsymbol{W}$ according to its selected order}}{
  Find the set of source nodes $\boldsymbol{S}$ which speak the word $w$. \\
  Initialize the set $\boldsymbol{H}$ of nodes, which hear $w$, to be null. \\
  \ForEach{{\rm node} $i \in \boldsymbol{U}$}{
   Determine whether $i$ can hear $w$ by $p_h$ (equation (6)). \\
   Add $i$ into $\boldsymbol{H}$ if $i$ heard $w$.}
  \ForEach{{\rm node} $i \in \boldsymbol{H}$}{
  If $w$ is in $i$'s memory, it is successful, delete $i$ from $\boldsymbol{U}$; otherwise, it is unsuccessful.}
  Count the number of successful nodes, $n_{succ}$. \\
  \ForEach{{\rm node} $i \in \boldsymbol{S}$}{
  If $i \in \boldsymbol{U}$, set it be successful with probability $n_{succ} / N$. \\
  Delete $i$ from $\boldsymbol{U}$ if $i$ becomes successful.}
 }
}
\end{algorithm*}

\section*{Appendix: The Model of Naming Game in Groups (NGG)}

Initially, every node (agent) in a given network has an empty memory and it is assumed that the memory of each node is infinite. During the NGG process, the rule for node $i$ to speak a word is as follows: if the memory of node $i$ is not empty, then it speaks a word randomly chosen from its memory; otherwise, it randomly chooses a word from a pre-set vocabulary $\boldsymbol{V}$. For any transmitting-word $w$, if node $i$ is successful then the node will clear out all other words from its memory but keeping only $w$; if node $i$ is unsuccessful then it will add the new word $w$ into its memory.

The model has several parameters: $M$ is the underlying network size; $N$ is the group size; $\beta$ is the proportion of transmitting-words with respect to $N$; $I_p$ is the pair-level weight for a pair of nodes, $I_n$ is the node-level weight, $I_w$ is the word-level weight, $p_w$ is the selecting probability for a word, $p_h$ is the hearing probability; $\boldsymbol{V}$ is a sufficiently large pre-set vocabulary; $\boldsymbol{A}$ is the adjacency matrix of the given network with $M$ nodes. The overall algorithm is summarized in Algorithm \ref{NGG}.

\section*{Acknowledgement}
The work described in this paper was fully supported by a grant from the Research Grants Council of the Hong Kong Special Administrative Region, China [Project No. CityU123312].


\begin{thebibliography}{}\label{sec:TeXbooks}
\bibitem{Axelrod81} Axelrod R, Hamilton WD: The evolution of cooperation. Science 211(4489): 1390-1396, 1981.
\bibitem{Barab99} Barab\'asi A-L, Albert R: Emergence of scaling in random networks. Science 286(5439): 509-512, 1999.
\bibitem{Baronchelli11} Baronchelli A: Role of feedback and broadcasting in the naming game. Phys Rev E 83: 046103, 2011.
\bibitem{Baronchelli07}	Baronchelli A, Dall'Asta L, Barrat A, Loreto V: The role of topology on the dynamics of the Naming Game. Eur Phys J Special Topics 143: 233–235, 2007.
\bibitem{Baronchelli06}	Baronchelli A, Felici M, Loreto V, Caglioti E, Steels L: Sharp transition towards shared vocabularies in multi-agent systems. J Stat Mech 2006: P06014, 2006.
\bibitem{Baronchelli10} Baronchelli A, Gong T, Puglisi A, Loreto V: Modeling the emergence of universality in color naming patterns. Proc Natl Acad Sci 107: 2403-2407, 2010.
\bibitem{Baronchelli08} Baronchelli A, Loreto V, Steels L: In-depth analysis of the naming game dynamics: The homogeneous mixing case. Int J Mod Phys C 19(05): 785-812, 2008.
\bibitem{Barrat07} Barrat A, Baronchelli A, Dall'Asta L, Loreto V: Agreement dynamics on interaction networks with diverse topologies. Chaos 17: 026111, 2007.
\bibitem{BOM14} Box Office Mojo. All time Box Office: Worldwide. http://www.boxofficemojo.com/alltime/world, 2014. Date of access: 25/07/2014.
\bibitem{Brigatti08} Brigatti E: Consequence of reputation in an open-ended naming game. Phys. Rev. E 78: 046108, 2008.
\bibitem{Castellano09} Castellano C, Fortunato S, Loreto V: Statistical physics of social dynamics. Rev Mod Phys 81: 591-646, 2009.
\bibitem{Conradt09} Conradt L, List C: Group decisions in humans and animals: a survey. Phil. Trans. R. Soc. B. 364: 719-742, 2009.
\bibitem{Dall06a} Dall'Asta L, Baronchelli A, Barrat A, Loreto V: Non-equilibrium dynamics of language games on complex networks. Phys Rev E 74: 036105, 2006.
\bibitem{Dall06b} Dall'Asta L, Baronchelli A, Barrat A, Loreto V: Agreement dynamics on small-world networks. Euro Phys Lett 73: 969, 2006.
\bibitem{Erd59} Erd\"os P, R\'enyi A: On random graphs I. Publ. Math. 6: 290-297, 1959.
\bibitem{Fu08} Fu F, Wang L: Coevolutionary dynamics of opinions and networks: From diversity to uniformity. Phys. Rev. E 78: 016104, 2008.
\bibitem{Li13} Li B, Chen G, Chow TWS: Naming game with multiple hearers. Commun Nonlinear Sci Numer Simulat 18: 1214-1228, 2013.
\bibitem{Liu09} Liu RR, Jia CX, Yang HX, Wang BH: Naming game on small-world networks with geographical effects. Physica A 388: 3615-3620, 2009.
\bibitem{Liu11} Liu RR, Wang WX, Lai YC, Chen G, Wang BH: Optimal convergence in naming game with geography-based negotiation on small-world networks. Phys Lett A 375: 363-367, 2011.
\bibitem{Liu13}	Liu Q, Wang, X: Opinion dynamics with similarity-based random neighbors. Sci Rep 3: 02968, 2013.
\bibitem{Loreto07} Loreto V, Steels L: Social dynamics: Emergence of language. Nat Phys 3(11): 758-760, 2007.
\bibitem{Lu09} Lu Q, Korniss G, Szymanski BK: The naming game in social networks: community formation and consensus engineering. J Economic Interaction Coordination 4(2): 221-235, 2009.
\bibitem{Maity12} Maity SK, Manoj TV, Mukherjee A: Opinion formation in time-varying social networks: The case of the naming game. Phys Rev E 86(3): 036110, 2012.
\bibitem{Maity13} Maity SK, Mukherjee A, Tria F, Loreto V: Emergence of fast agreement in an overhearing population: the case of naming game. Euro. Phys. Lett. 101: 68004, 2013.
\bibitem{Puglisi08} Puglisi A, Baronchelli A, Loreto V: Cultural route to the emergence of linguistic categories. Proc Natl Acad Sci 105: 7936-7940, 2008.
\bibitem{Rands03} Rands SA, Cowlishaw G, Pettifor RA, Rowcliffe JM, Johnstone RA: Spontaneous emergence of leaders and followers in foraging pairs. Nature 423, 432-434, 2003.
\bibitem{Steels96} Steels L: Self-organizing vocabularies. Artif Life 2: 319-332, 1996.
\bibitem{Steels12} Steels L, Loetzsch M: The grounded naming game. In: Experiments in Cultural Language Evolution (L Steels, ed.), John Benjamins Pub. Co., Amsterdam, 111-141, 2012.
\bibitem{Tang07} Tang CL, Lin BY, Wang WX, Hu MB, Wang BH: Role of connectivity-induced weighted words in language games. Phys Rev E 75: 027101, 2007.
\bibitem{Vylder06} Vylder BD, Tuyls K: How to reach linguistic consensus: A proof of convergence for the naming game. J Theor Biol 242(4): 818-831, 2006.
\bibitem{Wang07} Wang WX, Lin BY, Tang CL, Chen G: Agreement dynamics of finite-memory language games on networks. Eur Phys J B 60: 529-536, 2007.
\bibitem{Watts98} Watts DJ, Strogatz S: Collective dynamics of small-world networks. Nature 393: 440-442, 1998.
\bibitem{Xie11} Xie J, Sreenivasan S, Korniss G, Zhang W, Lim CC, Szymanski BK: Social consensus through the influence of committed minorities. Phys Rev E 84: 011130, 2011.
\bibitem{Yang08} Yang HX, Wang WX, Wang BH: Asymmetric negotiation in structured language games. Phys Rev E 77: 027103, 2008.
\bibitem{Zhang10} Zhang W, Lim CC: Noise in naming games, partial synchronization and community detection in social networks. arXiv:1008.4115, 2010.
\bibitem{Supp} Supplementary Metarial. \url{http://www.ee.cityu.edu.hk/~gchen/pdf/SM7.pdf}, 2014.







\end{thebibliography}
\end{document}